\documentclass[a4paper,11pt]{article}
\usepackage{pos}

\usepackage[utf8]{inputenc}
\usepackage{amsmath}
\usepackage{amsfonts}
\usepackage{subcaption}
\usepackage[numbers,sort&compress]{natbib}
\usepackage{url}
\usepackage{hyperref}
\usepackage{amsthm}
\usepackage{latexsym}

\usepackage{ascmac}
\usepackage{tikz}
\usepackage{here}
\usepackage{braket}
\usepackage{comment}
\usepackage[T1]{fontenc}
\usepackage{simplewick}
\usepackage{mathrsfs}
\usepackage{physics}
\usepackage{enumerate}
\usepackage{pgfplots}
\usepackage{cancel}
\usepackage{wrapfig}

\newcommand{\fix}[1]{\textcolor{red}{[#1]}}

\usepackage{pxpgfmark}

\usepackage{tikz}

\theoremstyle{definition} 

\newtheorem*{thm*}{Theorem}

\newtheorem*{defn*}{Definition}

\newtheorem*{lem*}{Lemma}

\newtheorem*{rem*}{Remark}

\newtheorem*{con*}{Conjecture}

\newtheorem*{cor*}{Corollary}

\newtheorem*{prop*}{Proposition}

\newtheorem*{hypoth*}{Hypothesis}

\newtheorem*{claim*}{Claim}

\makeatletter
\def\WordCount#1{%
	\@tempcnta\z@
	\@tfor \@tempa:=#1\do{\advance\@tempcnta\@ne}%
	\begin{quote}#1\end{quote}
	\the\@tempcnta 文字
}
\makeatother 

\def\rnum#1{\resizebox{0.5em}{\height}{\expandafter{\romannumeral #1}}}
\def\Rnum#1{\resizebox{0.5em}{\height}{\uppercase\expandafter{\romannumeral #1}}}

\newcommand{\Slash}[1]{{\ooalign{\hfil/\hfil\crcr\(#1\)}}}

\begin{document}
\title{Chiral fermion on curved domain-wall}
\author*[a]{Shoto Aoki}
\author[a]{Hidenori Fukaya}
\affiliation[a]{Department of Physics, Osaka University,\\Toyonaka, Osaka 560-0043, Japan}
\emailAdd{saoki@het.phys.sci.osaka-u.ac.jp}
\emailAdd{hfukaya@het.phys.sci.osaka-u.ac.jp}
\abstract{


We consider a massive fermion system having a curved domain-wall embedded in a square lattice. In a similar way to the conventional flat domain-wall fermion, chiral massless modes appear at the domain-wall but these modes feel "gravity" through the induced spin connections. In this work, we embed $S^1$ and $S^2$ domain-walls into a Euclidean space and show how the gravity is detected from the spectrum of the Dirac operator.
}

\FullConference{
The 38th International Symposium on Lattice Field Theory\\
Zoom/Gather@Massachusetts Institute of Technology\\
26th-30th July, 2021
}
\maketitle

\section{Introduction}

Formulating lattice field theory with a general curved metric has been a challenge, in contrast to the great success of gauge theory on a square lattice, of which continuum limit is a flat torus. In previous studies \cite{Hamber2009Quantum,Regge1961general,brower2016quantum,AMBJORN2001347Dynamicallytriangulating,Brower2017LatticeDirac}, they tried to use triangular lattices to represent dynamical/non-dynamical curved metric by changing the lengths or angles of the links.

In this work, we attempt a fermion formulation with a nontrivial spin connection but still staying on a square lattice. In mathematics, it is known that every curved compact manifold can be embedded into a higher-dimensional flat Euclidean space \cite{Nash1956TheImbedding}. If we regularize this higher-dimensional Euclidean space by a square lattice and localize a field on a curved sub-manifold embedded in it, the fermion field would feel gravity or curvature through the spin connection induced by the embedding. 

Such an embedding can be realized by the so-called domain-wall fermion formulation \cite{Jackiw1976Solitons,CALLAN1985427,KAPLAN1992342AMethod,Shamir1993Chiral}. Changing the sign of the mass on a co-dimension one sub-manifold, a massless fermion is localized on the domain-wall. In this work, we investigate a fermion system both in continuum and on a lattice when the domain-wall is nontrivially curved. In fact, in a similar lattice model of condensed matter physics, it has been reported that an effective spin connection is induced \cite{Lee2009Surface,Imura2012Spherical,Parente2011Spin,Takane2013UnifiedDescription}, in the edge modes appearing on a curved surface.

Specifically we embed a circle $S^1$ in $\mathbb{R}^2$ or 2-dimensional square lattice, as well as $S^2$ in $\mathbb{R}^3$ and its lattice counter part. We find that a nontrivial spin connection is induced, and it is reflected in the eigenvalue spectrum of the domain-wall Dirac operator. The analytic computation in continuum and numerical lattice results reasonably agree, which implies that a classical naive continuum limit of the higher dimensional square lattice is valid.

\section{$S^1$ domain-wall}\label{sec:S^1 in R^2}
In this section, we embed $S^1$ with radius $r_0$ as a domain wall into 2-dimensional Euclidean space $\mathbb{R}^2$. We study the spectrum of the Dirac operator focusing on the edge-localized states on the domain-wall. We show that the edge modes are massless and follow a Dirac equation with a nontrivial spin connection, which is effectively induced by the embedding.


\subsection{Continuum analysis}\label{subsec:S^1 conti}

First we consider a Hermitian Dirac operator 
\begin{align}
    H&=\sigma_3 \qty(\sigma_1 \pdv{}{x} +\sigma_2 \pdv{}{y} + M\epsilon) 
    =\mqty(M\epsilon & e^{-i\theta}(\pdv{}{r}-\frac{i}{r}\pdv{}{\theta}) \\
    -e^{i\theta}(\pdv{}{r}+\frac{i}{r}\pdv{}{\theta})& -M\epsilon
    ) \label{eq:Dirac op S^1 conti},
\end{align}
in continuum theory, where an $S^1$ domain-wall with radius $r_0$ is put by the sign function $\epsilon= \text{sign}\qty( r-r_0)$. Since this Dirac operator commutes with the total angular momentum operator $J=-i\pdv{}{\theta}+\frac{1}{2}\sigma_3$, the solutions to the Dirac equation are the simultaneous eigenstates of $H$ and $J$\footnote{In the relativistic theory, where we treat $\mathbb{R}^2$ as a space-time, $H$ is not a Hamiltonian and $J$ is not an angular momentum in the spatial directions. However in our polar coordinate notation,  the spatial notation is easier to understand. Here and in the following, we, therefore,  call the eigenvalue of $H$ "energy", that of $J$ "angular momentum" and that of $\gamma_{\text{normal}}$ given in Eq. \eqref{eq:chirality op S^1}  "chirality". }. Below, let energy $E$ be an eigenvalue of $H$ and we represent that $j=\pm\frac{1}{2},\pm \frac{3}{2},\cdots$ are eigenvalues of $J$. Note that the energy $E$ of edge states must satisfy $E^2<M^2$.




In the interior of the domain-wall ($r<r_0$), the eigenfunction with the energy $E$ and total angular momentum $j=\pm\frac{1}{2},\pm \frac{3}{2},\cdots$ is written by
\begin{align}\label{eq:eigenfunc of edgemode in of S^1}
\begin{aligned}
    \psi^{E,j}_{r<r_0}&=\mqty(\sqrt{M^2-E^2} I_{j-\frac{1}{2}} (\sqrt{M^2-E^2} r)e^{(j-\frac{1}{2})i\theta }\\ 
         (M+E) I_{j+\frac{1}{2}} (\sqrt{M^2-E^2} r)e^{(j+\frac{1}{2})i\theta }
         ) 
\end{aligned}.
\end{align}
Similarly, the eigenfunction in the exterior $r>r_0$ with the energy $E=\pm \abs{E}\ (0<\abs{E}<M)$ and total angular momentum $j$ is given by
\begin{align}\label{eq:eigenfunc of edgemode out of S^1}
\begin{aligned}
        \psi^{E,j}_{r>r_0}&=\mqty((M+E)K_{j-\frac{1}{2}} (\sqrt{M^2-E^2} r)e^{(j-\frac{1}{2})i\theta }\\ 
         \sqrt{M^2-E^2} K_{j+\frac{1}{2}} (\sqrt{M^2-E^2} r)e^{(j+\frac{1}{2})i\theta }
         )
\end{aligned}.
\end{align}
The eigenfunctions Eqs. \eqref{eq:eigenfunc of edgemode in of S^1} and \eqref{eq:eigenfunc of edgemode out of S^1} are exponentially localized at $r=r_0$. From the continuity requirement of the wave functions at $r=r_0$, we have a condition
\begin{align}\label{eq:condition of E}
    \frac{I_{j-\frac{1}{2}}}{I_{j+\frac{1}{2}}}\frac{K_{j+\frac{1}{2}}}{K_{j-\frac{1}{2}}}(\sqrt{M^2-E^2}r_0)=\frac{M+E}{M-E}.
\end{align}

In the large mass limit or $M\gg E$, the energy eigenvalue converges to
\begin{align}
    E\simeq \frac{j}{r_0}\ \qty(j=\pm\frac{1}{2},\pm\frac{3}{2},\cdots). 
\end{align}
where we can recognize the gap from zero as a gravitational effect on
the curved domain-wall.

The normalized eigenfunction in that limit is also simplified as\footnote{
This approximation is only valid at $r\sim r_0$.
}
\begin{align}\label{eq:asymptotic form of edgemode S^1}
    \psi_\text{edge}^{E,j}\simeq \sqrt{\frac{M}{4\pi r}}e^{-M\abs{r-r_0}}\mqty( e^{i(j-\frac{1}{2})\theta}\\ e^{i(j+\frac{1}{2})\theta}), 
\end{align}
which is chiral with respect to a gamma matrix facing the normal direction to the domain-wall,
\begin{align}\label{eq:chirality op S^1}
    \gamma_{\text{normal}}:=&\sigma_1 \cos\theta+ \sigma_2 \sin \theta, 
\end{align}
with the eigenvalue $+1$. Since the edge modes are localized at the $S^1$ domain-wall, it is natural to assume that the effective action of the edge modes is written as that of a one-dimensional Dirac fermion. To confirm this, let us  take a linear combination of the edge modes by
\begin{align}
    \psi_\text{edge} &= \sqrt{\frac{M}{4\pi r}}e^{-M\abs{r-r_0}}\mqty(
1 \\ e^{i \theta}) \chi(\theta),\\   \chi(\theta)& = \sum_j \alpha_j e^{i (j-\frac{1}{2})\theta},
\end{align}
where $\alpha_i$ is a complex number coefficient. The effective action is then obtained as
\begin{align}
    \lim_{ M\to \infty}\int dx^1 dx^2 \psi^\dagger_{\text{edge}}H\psi_{\text{edge}} 
    = \int_0^{2\pi} d\theta \chi^\dagger\frac{1}{r_0}\qty( -i\pdv{}{\theta}+\frac{1}{2})\chi \label{eq:effective op S^1 spinc} ,
\end{align}
where the effective Dirac operator can be read as
$J=-i\pdv{}{\theta}+\frac{1}{2}\sigma_3$ for states with $\sigma_3=1$.

In Eq. \eqref{eq:effective op S^1 spinc}, the covariant derivative contains a nontrivial gauge potential
$1/2r_0$, which can be regarded as the induced spin connection\footnote{More precisely, this term is a $\text{spin}^c$ connection rather than a spin connection.} field.
It is also clear that this connection is the origin of the energy gap from zero.
Namely, the edge localized fermion feels induced gravity from the
curved domain-wall.
Since the gravity in one dimension is locally trivial, we can absorb
the effect by a gauge transformation $\chi \to
\chi^\prime=\exp(-i\theta/2) \chi$
and action becomes that of a free Dirac fermion,
\begin{align}
    \int_0^{2\pi} d\theta \chi^\dagger\frac{1}{r_0}\qty( -i\pdv{}{\theta}+\frac{1}{2})\chi = 
    \int_0^{2\pi} d\theta (\chi^\prime)^\dagger\frac{1}{r_0}\qty( -i\pdv{}{\theta})\chi^\prime .
\end{align}
The global effect of gravity is still visible as the anti-periodicity
of $\chi^\prime$,
which makes the same energy gap as the one with $\chi$.

\subsection{Lattice analysis}\label{subsec:S^1 lattice}
Let $(\mathbb{Z}/n\mathbb{Z})^2$ be a 2-dimensional square lattice space. $0\leq x,y \leq n-1 $ denotes coordinates for $(\mathbb{Z}/n\mathbb{Z})^2$, where $n$ means the lattice size. We impose the periodic boundary condition in the $x$ and $y$ directions as
\begin{align}
    x=0\sim n,\ y=0\sim n.
\end{align}

On this lattice, we consider a Wilson Dirac operator with a domain-wall mass, 
        \begin{align}
        H =\sigma_3 \qty(\sum_{i=1,2}\qty[\sigma_i\frac{\nabla^f_i+\nabla^b_i}{2} -\frac{r}{2}\nabla^f_i \nabla^b_i ]+\epsilon_A M ) \label{eq:Hermitian Wilson Dirac op}\\
        (\nabla^f_i \psi)_x =\psi_{x+\hat{i}}-\psi_{x},\ (\nabla^b_i \psi)_x =\psi_{x}-\psi_{x-\hat{i}}. \nonumber
    \end{align}
Here, we assign the domain-wall mass by a step function
\begin{align}
    \epsilon_A(x)=\left\{ \begin{array}{cc}
        -1 & (x\in A)  \\
        1 & (x\notin A)
    \end{array}\right. ,
\end{align}
where the region $A$ inside the circle is defined by
\begin{align}
    A=\Set{(x,y)\in (\mathbb{Z}/n\mathbb{Z})^2 | \qty(x-\frac{n-1}{2})^2+\qty(y-\frac{n-1}{2})^2< (r_0)^2},
\end{align}
where $r_0$ is the radius of $S^1$ domain-wall in the lattice units. 

\begin{figure}[h]
  \begin{minipage}{0.45\linewidth}
    \centering
    \includegraphics[scale=0.45,bb= 0 0 461 346]{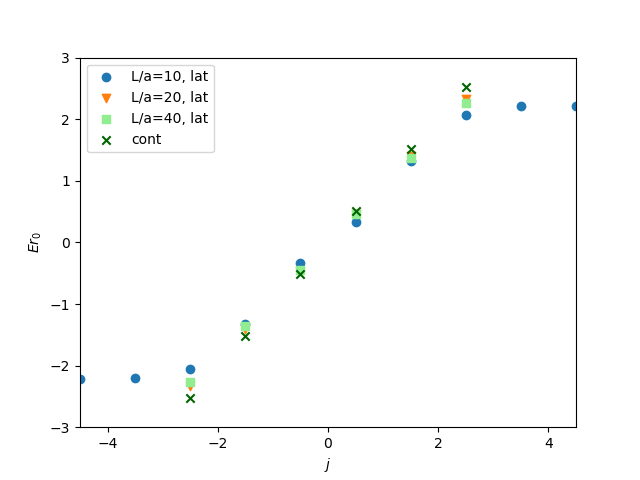}
    \caption{The Dirac eigenvalue spectrum normalized by the circle radius at $Ma=0.7$, $r_0=L/4$.}
    \label{fig:Spectrum S^1 lat}
  \end{minipage}
  \hfill
  \begin{minipage}{0.45\linewidth}
    \centering
    \includegraphics[scale=0.45,bb= 0 0 461 346]{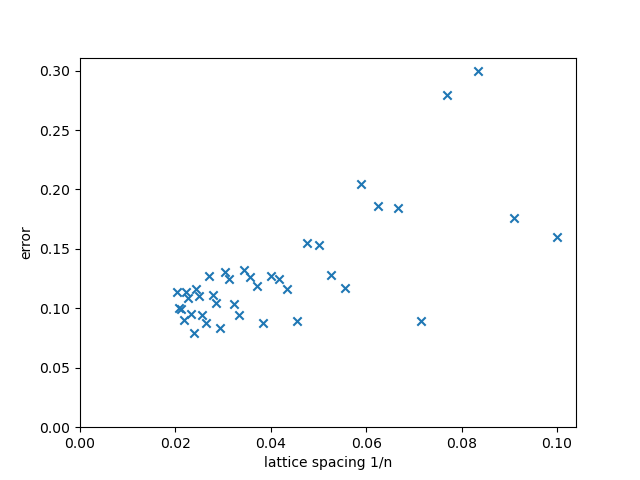}
    \caption{
    The relative deviation of the lowest eigenvalue
${\abs{E_{\frac{1}{2}}^{\text{con}}-E_{\frac{1}{2}}^\text{lat}}}/{E_{\frac{1}{2}}^{\text{con}}}$
 at $r_0=\frac{L}{4}$ is plotted as a function of the lattice spacing $1/n$.
    }
    \label{fig:error_S1}
  \end{minipage}
\end{figure}

We solve the eigenvalue problem of $H$ numerically, and plot the eigenvalues normalized by the radius $r_0$ in Fig. \ref{fig:Spectrum S^1 lat}. Here we take $Ma=0.7$ and $r_0=L/4$. Data at different lattice spacings $n=L/a =10,20, 40$ are plotted with circles, triangles and squares, respectively. Here we label the eigenvalues with half integer $j$, expecting them to  represent the eigenstates of $J$ in the continuum limit,
\begin{align}
    \cdots\leq E_{-\frac{3}{2}}\leq E_{-\frac{1}{2}} \leq 0 \leq E_{\frac{1}{2}} \leq E_{\frac{3}{2}} \leq \cdots.
\end{align}




In fact, the lattice data look converging to the continuum limit denoted by cross symbols in Fig.~\ref{fig:Spectrum S^1 lat}. The relative deviation of $E_{\frac{1}{2}}$ from the continuum limit as a function of the lattice spacing is plotted in Fig.~\ref{fig:error_S1}. Although it is not monotonic due to the violation of the rotational symmetry, the deviation decreases as $n$ grows.

\begin{figure}
                    \centering
             \includegraphics[width=0.45\textwidth,bb=0 0 389 373]{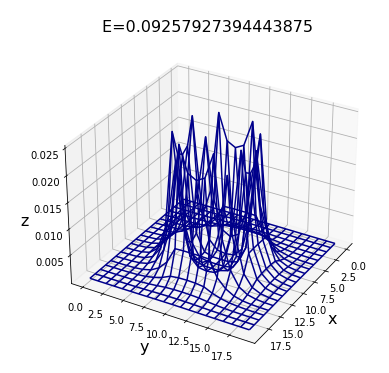}
              \caption{The amplitude of the eigenfunction with $E_{\frac{1}{2}}$.}
              \label{fig:S^1 edge state on lattice}
          \end{figure}

In Fig.~\ref{fig:S^1 edge state on lattice}, we present the distribution of the eigenfunction amplitude. As is expected, the wave function is localized at the domain-wall. Moreover, we confirm that the state is chiral : the expectation value of $\gamma_\text{normal} = 0.99 $. We conclude that the massless fermion on $S^1$, feeling gravity through the induced spin connection, can be formulated on a two-dimensional square lattice by the curved domain-wall fermion.



\section{$S^2$ domain-wall}\label{sec:S^2 in R^3}
In this section, we embed $S^2$ with a radius $r_0$ as a domain wall into 3-dimensional Euclidean space $\mathbb{R}^3$. 

\subsection{Continuum analysis}
As in the previous section, we consider a Hermitian Dirac operator 
\begin{align}
    H= \gamma^0 \qty(\gamma^j \pdv{}{x^j}+M\epsilon )=\mqty(M \epsilon & \sigma^j \partial_j \\ -\sigma^j \partial_j & -M\epsilon ),\ 
    ({\gamma}^0=\sigma_3 \otimes 1,\ {\gamma}^j=\sigma_1\otimes \sigma_j) 
\end{align}
where an $S^2$ domain-wall with radius $r_0$ is put by the sign function $\epsilon= \text{sign}\qty( r-r_0)$. Note that this operator acts on $4$-component spinors. This operator commutes with the total angular momentum and parity operators given by  
\begin{align}
    J_i&=1\otimes \hat{J}_i=1\otimes (L_i+\frac{1}{2}\sigma_i),\\
    P\psi(x)&=(\sigma_3\otimes 1) \psi(-x),
\end{align}
where $\hat{J}_i$ acts on $2$-component spinors. The eigenstates of the Dirac operator are the simultaneous eigenstates of $H,J^2=(J_1)^2+(J_2)^2+(J_3)^2,J_3$ and $P$. Let energy $E$ be an eigenvalue of $H$ and we represent that $j(j+1)\ (j=\frac{1}{2},\frac{3}{2},\cdots)$ are eigenvalues of $J^2$ and $j_3=-j,-j+1,\cdots,j$ are eigenvalues of $J_3$ for the states with $J^2=j(j+1)$. Note that the energy $E$ of edge states must satisfy $E^2<M^2$.


We solve the eigenvalue problem of $H$ in a similar way to the previous section. At first, we introduce a 2-component spinor $\chi_{j,j_3}^{(\pm)}$ that satisfies the following equations:
\begin{align}
    \hat{J}^2\chi_{j,j_3}^{(\pm)}&=j(j+1)\chi_{j,j_3} ^{(\pm)}\\
    \hat{J}_3  \chi_{j,j_3}^{(\pm)}&= j_3 \chi_{j,j_3}^{(\pm)} \\
    \chi_{j,j_3}^{(\pm)}(-x)&=(-1)^{j\mp \frac{1}{2}}\chi_{j,j_3}^{(\pm)}(x) \\
    \chi_{j,j_3}^{(-)}&=\frac{\sigma\cdot x}{r}\chi_{j,j_3}^{(+)}.
\end{align}
With these states, we find eigenstates with energy $E>0 $, total angular momentum $j$, and $j_3$ as
\begin{align}\label{eq:edgestate S^2 E>0}
    \psi^{E>0}_{j,j_3,+}&=\left\{
\begin{array}{ll}
     \frac{1}{\sqrt{r}}\mqty( \sqrt{M^2-E^2} I_{j }(\sqrt{M^2-E^2}r) \chi_{j,j_3}^{(+)} \\(M+E) I_{j+1}(\sqrt{M^2-E^2}r) \frac{\sigma\cdot x}{r}\chi_{j,j_3}^{(+)} )  & (r<r_0) \\
     \frac{c}{\sqrt{r}}\mqty( (M+E) K_{j }(\sqrt{M^2-E^2}r)\chi_{j,j_3}^{(+)} \\\sqrt{M^2-E^2} K_{j+1}(\sqrt{M^2-E^2}r) \frac{\sigma\cdot x}{r}\chi_{j,j_3}^{(+)} ) & (r>r_0)
\end{array}
\right.
\end{align}
and eigenstates with energy $E<0 $, total angular momentum $j$, and $j_3$ as
\begin{align}\label{eq:edgestate S^2 E<0}
    \psi^{E<0}_{j,j_3,-}&=\left\{
\begin{array}{ll}
\frac{1}{\sqrt{r}}\mqty( (M-E) I_{j+1}(\sqrt{M^2-E^2}r) \chi_{j,j_3}^{(-)} \\\sqrt{M^2-E^2} I_{j}(\sqrt{M^2-E^2}r) \frac{\sigma\cdot x}{r}\chi_{j,j_3}^{(-)} ) & (r<r_0) \\
     \frac{c^\prime}{\sqrt{r}}\mqty( \sqrt{M^2-E^2} K_{j+1 }(\sqrt{M^2-E^2}r) \chi_{j,j_3}^{(-)} \\(M-E) K_{j}(\sqrt{M^2-E^2}r) \frac{\sigma\cdot x}{r}\chi_{j,j_3}^{(-)} ) & (r>r_0)
\end{array}
\right. ,
\end{align}
where coefficients $c,c^\prime$ are numerical constants. From the continuous condition at $r=r_0$, $\abs{E}$ must satisfy
\begin{align}\label{eq:condition of E S^2}
    \frac{I_{j}}{I_{j+1}}\frac{K_{j+1}}{K_{j}}(\sqrt{M^2-\abs{E}^2}r_0)=\frac{M+\abs{E}}{M-\abs{E}}.
\end{align}
Note that the eigenfunctions have $2j+1$-fold degeneracy with different $j_3$. In the large mass limit or $M\gg E$, the energy eigenvalues converges to 
\begin{align}
    \abs{E}\simeq \frac{j+\frac{1}{2}}{r_0},\ \qty(j=\frac{1}{2}, \frac{3}{2}\cdots).
\end{align}
Since $j$ is positive, the spectrum of $H$ has a gap from zero, which is again a gravitational footprint made by the embedding of the curved domain-wall.

The normalized edge modes $\tilde{\psi}^{E=\pm \abs{E}}_{j,j_3,\pm}$ is also simplified as
\begin{align}
    \tilde{\psi}^{E=\pm \abs{E}}_{j,j_3,\pm}\simeq \sqrt{\frac{M}{2}}\frac{e^{-M\abs{r-r_0}}}{r}\mqty(\chi_{j,j_3}^{(\pm)} \\ \frac{\sigma \cdot x}{r}\chi_{j,j_3}^{(\pm)}). 
\end{align}
This state is a eigenstate of a gamma matrix facing the normal direction of the $S^2$ domain-wall,
    \begin{align}
        \gamma_{\text{normal}}:=&\sum_{i=1}^3 \frac{x^i}{r} {\gamma}^i
    \end{align}
with eigenvalue $+1$.


In the same way as the previous section, we can define the effective two-component spinor on $S^2$ by a linear combination $\chi=\sum a_{j,j_3}^{(\pm)}\chi_{j,j_3}^{(\pm)}$ and we obtain the effective action
\begin{align}
    \lim_{M\to \infty}\int_{\mathbb{R}^3} d^3x (\psi_\text{edge})^\dagger H\psi_{\text{edge}} 
    =\int_{S^2} dS \chi^\dagger\frac{1}{r_0}(\sigma\cdot L+1) \chi.
\end{align}

With the gauge transformation
\begin{align}
    s^{-1}=\mqty(e^{i\frac{\phi}{2}} \cos \frac{\theta}{2} & e^{-i\frac{\phi}{2}} \sin\frac{\theta}{2} \\-e^{i\frac{\phi}{2}} \sin\frac{\theta}{2} &e^{-i\frac{\phi}{2}} \cos \frac{\theta}{2}
    ),
\end{align}
$H_{S^2}:=\frac{1}{r_0}(\sigma\cdot L+1)$ is transformed as
\begin{align}
    s^{-1} H_{S^2} s&=-\frac{\sigma_3}{r_0} \qty(\sigma_1 \pdv{}{\theta} +\sigma_2 \qty( \frac{1}{\sin \theta} \pdv{}{\phi}  -\frac{\cos\theta}{2 \sin\theta} \sigma_1 \sigma_2 ) ) =-\frac{\sigma_3}{r_0} \Slash{D}_{S^2},
\end{align}
where we find again a nontrivial spin connection $-\frac{\cos\theta}{2\sin\theta} \sigma_1 \sigma_2$ on $S^2$. The induction of the nontrivial connection was also reported in condensed matter physics
\cite{Imura2012Spherical,Parente2011Spin,Takane2013UnifiedDescription}.

\subsection{Lattice analysis}

Let $(\mathbb{Z}/n\mathbb{Z})^3$ be a 3-dimensional square lattice space. $0\leq x,y,z \leq n-1 $ denote coordinates for $(\mathbb{Z}/n\mathbb{Z})^3$, where $n$ means a length of one edge of the lattice space. We impose periodic boundary condition in every direction.

We consider a Hermitian Wilson-Dirac operator 
    \begin{align}
        H =\gamma_0 \qty(\sum_{i=1,2,3}\qty[\gamma_i\frac{\nabla^f_i+\nabla^b_i}{2} -\frac{r}{2}\nabla^f_i \nabla^b_i ]+\epsilon M ), \label{eq:Hermitian Wilson Dirac op}\\
        (\nabla^f_i \psi)_x =\psi_{x+\hat{i}}-\psi_{x},\ (\nabla^b_i \psi)_x =\psi_{x}-\psi_{x-\hat{i}}, \nonumber
    \end{align}
where the step function $\epsilon=\text{sign}(r-r_0)$ represents the $S^2$ domain-wall. 

\begin{figure}
    \centering
    \includegraphics[scale=0.4,bb=0 0 461 346]{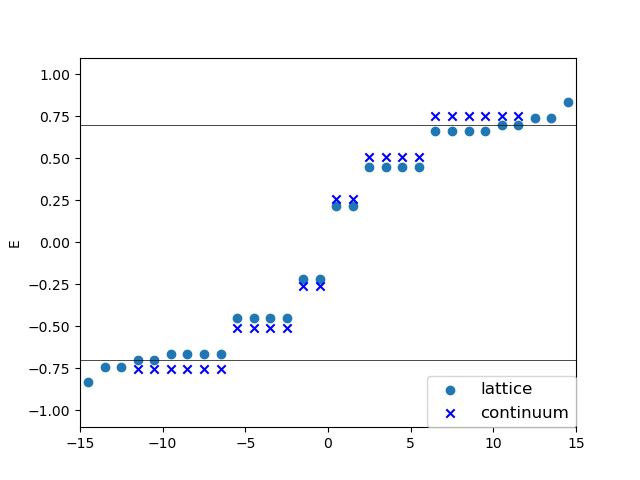}
    \caption{The eigenvalues of the edge states on the $S^2$ domain-wall at $n=16,M=0.7$.
The lattice data are plotted by circles, while the continuum limits
are presented by crosses.}
    \label{fig:Spectrum of edge states at S^2}
\end{figure}

\begin{figure}
  \begin{minipage}[b]{0.45\linewidth}
    \centering
            \includegraphics[width=\textwidth,bb= 0 0 372 299]{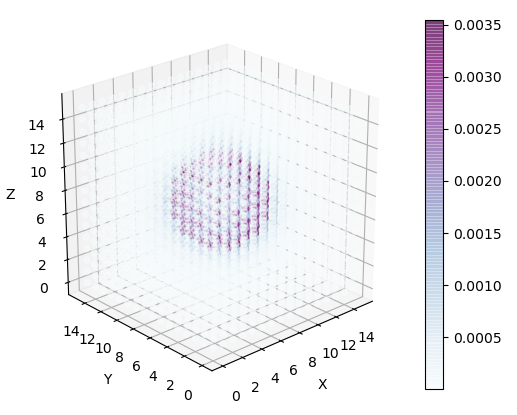}
    \subcaption{Overall picture.}
  \end{minipage}
  \begin{minipage}[b]{0.45\linewidth}
    \centering 
    \includegraphics[width=0.8\textwidth,bb=0 0 315 307]{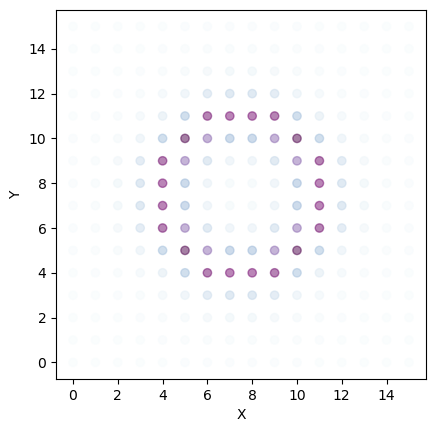}
    \subcaption{Slice at $z=7$.}
  \end{minipage}
  \caption{The amplitude of the eigenfunction with $E=0.218\cdots$ at $M=0.7$ and lattice
size $=16^3$.}
  \label{fig:edge state of S^2}
\end{figure}

Solving an eigenvalue problem of this operator numerically and arranging the eigenvalues in descending order, we obtain  the spectrum presented by circle symbols in Fig.~\ref{fig:Spectrum of edge states at S^2}. We find a reasonable agreement with that in continuum shown by crosses, including the gap from zero. As shown in Fig.~\ref{fig:edge state of S^2}, we find that the eigenstates with the energy $\abs{E}<M$ are localized at the $S^2$ domain-wall, which is again a good evidence for the success of formulating the fermion system with a nontrivial curved background.






\section{Conclusion}\label{sec:Conclusion}
%

In this work, we have considered $S^1$ and $S^2$ curved backgrounds embedded as domain-walls into a higher-dimensional Euclidean square lattice. We have found that the chiral localized states appear at the domain-wall. We have also found numerical evidences that these edge states feel gravity through the induced spin connection. The reasonable agreement with the continuum limits is encouraging in that a naive continuum limit of the higher-dimensional square lattice is valid.


\acknowledgments 
We thank M. Furuta, K. Hashimoto, M. Kawahira, S. Matsuo, T. Onogi, S. Yamaguchi and M. Yamashita for useful discussions. This work was supported in part by JSPS KAKENHI Grant Number
JP18H01216 and JP18H04484.


\bibliographystyle{JHEP}
\bibliography{ref}

\end{document}